\documentclass{article}
\usepackage{graphicx} 
\usepackage{url}
\usepackage{amsmath}
\usepackage{booktabs}
\usepackage{float}
\usepackage{array}
\usepackage{makecell}
\usepackage{caption}
\usepackage{multirow}
\usepackage{changepage} 
\usepackage[margin=1.25in]{geometry} 
\usepackage{authblk}
\usepackage{algorithm}
\usepackage{algpseudocode}

\usepackage[
    backend=biber,
    style=numeric,
    sorting=none,
    maxbibnames=4,
    minbibnames=4,
    maxcitenames=2,
    mincitenames=1,
    giveninits=true
]{biblatex}

\title{Testing LLM Arithmetic Reasoning Generalization with Automatic Numeric-Remapping Attacks}

\author[1]{Malia Barker}
\author[1]{Bishal Lakha}
\author[1]{Edoardo Serra}
\author[2]{Francesco Gullo}

\affil[1]{Department of Computer Science, Boise State University, Boise, ID, USA}
\affil[2]{University of L'Aquila, L'Aquila, Italy}

\date{}

\begin{document}

\maketitle

\begin{abstract}
    Large language models achieve strong performance on arithmetic reasoning benchmarks, and one response to arithmetic brittleness is to delegate computation to code. However, models are still often used in settings where they must reason directly from natural language, and a trustworthy model should be able to solve small-number arithmetic word problems without relying on external tools. Recent work shows that LLMs are sensitive to numerical variation in math word problems. Models may solve an original problem correctly yet fail on structurally similar variants that require the same reasoning procedure but use different numerical values. Evidence from controlled numerical perturbations suggests that this fragility is driven substantially by brittle arithmetic execution, especially as numerical values grow larger, while numerical changes can also induce errors in reasoning steps. We ask whether this fragility persists under a stricter and more controlled setting: small, schema-preserving numeric changes that retain the original reasoning program and avoid large-number stress tests.

We introduce an algorithm able to generate numeric-remapping generalization attacks for arithmetic word problems. Unlike related template-based perturbation approaches that require manually constructed schemas, human-defined constraints, or substantial manual validation, our attack automatically derives problem-specific symbolic representations from individual word problems, generates constrained numeric remappings, and recomputes the attacked gold answer without manual annotation. The resulting pipeline extracts symbolic problem representations, realizes transformed questions through LLM-generated edit plans applied deterministically, and filters retained examples through stage-wise validation and a post-hoc high-confidence audit. This makes numeric-remapping attacks a scalable evaluation procedure for probing arithmetic robustness with limited human intervention.

We evaluate DeepSeek-R1 (70B), Gemma4 (31B), and GPT-OSS (120B) on high-confidence numeric-remapping attacks across GSM8K, MAWPS, and MultiArith. On GSM8K, completed evaluations show meaningful conditional accuracy drops, ranging from (12.16) to (25.82) percentage points across completed model runs. In contrast, MAWPS and MultiArith show much smaller drops, with most completed evaluations remaining near or above \(98\%\) attacked accuracy. These results show that numeric-remapping robustness depends strongly on dataset structure: GSM8K remains sensitive to controlled numeric variation even when transformed problems preserve the original reasoning program and use recomputed gold answers, while shorter and more regular datasets are more stable under the retained transformations. This suggests that benchmark performance, especially on GSM8K, can overstate the reliability of direct arithmetic reasoning and that numeric remapping provides a practical diagnostic for evaluating whether models have learned stable problem schemas or remain sensitive to the particular numerical instantiations present in fixed benchmarks.
\end{abstract}

\section{Introduction}
Despite impressive performance on standardized benchmarks, large language models remain surprisingly brittle. A model that performs well on a fixed benchmark may fail when the same underlying task is presented with small, meaning-preserving changes, such as a rephrased question, a renamed variable, or a modified surface form \cite{Sun2024TrustLLMTI,Kostic2026SameMD}. This instability has been observed across code generation, mathematical reasoning, logical inference, and commonsense tasks, with performance drops varying substantially by domain and model \cite{Hossen2024OnTA,Khurana2024AHL,Tian2025EvolProverAA}. These failures are especially concerning because they suggest that strong benchmark performance does not necessarily imply robust reasoning. A model may learn to exploit dataset-specific language, common templates, or familiar numeric patterns rather than acquiring a stable procedure that transfers across nearby instances of the same problem \cite{CohenInger2025ForgetWY,Lunardi2025OnRA}.

This issue is particularly important for arithmetic word problems. Benchmarks such as GSM8K are often used as evidence of mathematical reasoning ability, yet many arithmetic problems contain recurring schemas, conventional phrasings, and familiar quantitative relationships. Although models can sometimes use external tools or generate code to solve arithmetic tasks, direct natural-language arithmetic remains important: a trustworthy model should be able to reason through small-number word problems when tool use is unavailable, inappropriate, or not part of the evaluation setting. A model can answer the original version of a problem correctly while still failing when the concrete numbers are changed and the same reasoning structure must be applied again. Prior work on mathematical robustness, including GSM-Symbolic, GSM-Plus, numerical-variation benchmarks, and related perturbation-based evaluations, shows that controlled variants of math problems can reveal failures hidden by static benchmark accuracy \cite{Mirzadeh2024GSMSymbolicUT,li2024gsmplus,yang-etal-2025-evaluating,huang2025mathperturb}. These findings motivate evaluation methods that test not only whether a model solves a benchmark item once, but whether it remains correct under controlled, schema-preserving transformations.

In this paper, we study \textit{generalization attacks} for arithmetic word problems, focusing on numeric remapping. A numeric-remapping attack changes the concrete quantities in a problem while preserving the underlying reasoning program. For example, if an original problem requires adding a base quantity to half of that quantity, the attacked version may replace the base value and the fractional relation while keeping the same symbolic computation. The attacked answer is then recomputed from the transformed symbolic representation rather than produced heuristically. This allows the attacked example to remain a valid, standalone word problem with a verified gold answer.

We introduce a structured pipeline for generating numeric-remapping attacks. Starting from an original question-answer pair, the pipeline infers a symbolic representation of the arithmetic computation, extracts local constraints on editable quantities, proposes new values under those constraints, renders the transformed problem back into natural language, and recomputes the attacked gold answer. At each stage, intermediate artifacts are checked before being passed forward. This staged design is intended to reduce invalid transformations and ensure that retained attacks preserve the intended problem schema while changing the numeric instantiation.

We evaluate this pipeline on arithmetic word-problem benchmarks using large language models as answer-generation systems. The results show that numeric remapping can substantially reduce model performance on GSM8K, even when attacks are generated only from examples that the source model originally answered correctly, while smaller drops on MAWPS and MultiArith suggest that robustness varies by dataset structure and problem complexity. These results indicate that benchmark accuracy on original questions can overstate the stability of arithmetic reasoning under controlled numeric variation.

The contributions of this paper are as follows:

\begin{itemize}
    \item We define numeric remapping as a schema-preserving generalization attack for arithmetic word problems, in which concrete quantities are changed while the underlying reasoning program is preserved and the gold answer is recomputed.

    \item We present a structured attack-generation pipeline that extracts symbolic problem representations, generates constrained numeric transformations, renders attacked questions, and validates retained examples through stage-wise checks.

    \item We empirically evaluate large language models on numeric-remapping attacks across arithmetic word-problem datasets, showing that models that perform well on original benchmark questions can still degrade substantially on valid, recomputed variants.
\end{itemize}

\section{Related Work}
\subsection{Reasoning benchmarks and arithmetic word problems}
Arithmetic word problem benchmarks have evolved considerably in scope and complexity, spanning from elementary single-step problems to competition-level mathematics. The most widely adopted benchmark, GSM8K, contains 8,500 linguistically diverse grade-school math problems requiring 2–8 reasoning steps \cite{cobbe2021training}, while SVAMP \cite{Patel2021AreNM} is specifically designed to expose shallow heuristics through careful structural variations. For targeted arithmetic capabilities, AddSub\cite{Hosseini2014LearningTS} and MultiArith \cite{Roy2016SolvingGA} address multi-step arithmetic reasoning, ASDiv \cite{Miao2020ADC} emphasizes linguistic diversity across elementary problem types, and AQuA-RAT \cite{Ling2017ProgramIB} extends to algebraic reasoning with over 100K problems paired with natural language rationales. MAWPS \cite{KoncelKedziorski2015ParsingAW} consolidates many of these into a unified repository of 3,320+ problems, and more recently TabMWP \cite{Lu2022DynamicPL} pushes further by requiring joint reasoning over both textual and tabular data across 38,431 problems. As LLMs began saturating standard benchmarks, the community shifted toward more challenging datasets such as MATH \cite{Hendrycks2021MeasuringMP} with 12,500 competition-level problems, and OlympiadBench \cite{He2024OlympiadBenchAC} covering Olympiad-level mathematics and physics. 

\subsection{Robustness and distribution shift in LLMs}
Strong benchmark performance in LLMs does not reliably translate to robustness under real-world conditions. A growing body of evidence shows that even state-of-the-art models such as GPT-4-turbo, GPT-3.5-turbo, and Gemini-1.0-pro suffer performance drops exceeding 25\% when faced with knowledge-invariant perturbations of benchmark questions, with GPT-4-turbo achieving only 55.5\% consistency \cite{Li2024PertEvalUR}. This disconnect arises because high-scoring models tend to exploit dataset-specific surface cues rather than develop genuine language understanding, and when these cues are perturbed, performance collapses, revealing a fundamental paradox between benchmark accuracy and true generalization \cite{CohenInger2025ForgetWY}. The sensitivity extends across perturbation types: lexical modifications cause statistically significant degradation across nearly all models and tasks, while small phrasing changes can swing accuracy by as much as -42.1\% in one direction and +35.3\% in another, exposing both fragility and unpredictability in current reasoning systems \cite{Roh2025ChainofCodeCR}. Critically, benchmark rankings themselves are unstable, as minor changes such as reordering answer choices on MMLU can shift model rankings by up to 8 positions \cite{Alzahrani2024WhenBA}, undermining the validity of leaderboard-driven model selection.

The root cause lies in how LLMs are trained and evaluated. Standard pre-training via next-token prediction and fine-tuning via Empirical Risk Minimization optimize for average performance on the training distribution, without explicitly encouraging robustness to distributional shift or perturbation \cite{Kumar2025RobustnessIL}. This produces brittle solutions that degrade when deployment conditions deviate from curated benchmark settings \cite{Li2024PertEvalUR}, further compounded by data contamination that inflates apparent benchmark performance through memorization rather than reasoning \cite{He2023UsingNL}. In out-of-distribution settings, fine-tuned domain-specific models that outperform LLMs on in-distribution examples lose their advantage, with in-context learning proving more resilient for out-of-distribution (OOD) instances \cite{Yuan2023RevisitingOR}. Synthetic training data exacerbates this further, as uniform formatting induces pattern overfitting and output distribution shift, degrading instruction-following capabilities \cite{Chen2025StandardBF}. Collectively, these findings suggest that static benchmark scores are insufficient proxies for model reliability, and robustness to distribution shift remains critically under-tested, appearing in only 10 out of 19 supervised LLM uncertainty quantification methods surveyed \cite{Devic2025FromCT}.

\subsection{Adversarial attacks and perturbation-based evaluation}
As language models achieve stronger results on standard benchmarks, a growing line of work has begun to ask whether those results remain stable under small, controlled changes \cite{wang2022advglue,Li2024PertEvalUR}. Perturbation-based evaluation operates by taking an existing benchmark example, introducing a controlled change to its wording, structure, numerical values, or presentation, and then measuring whether the model continues to solve the problem correctly. Rather than simply making the task more difficult, this framework is intended to assess whether benchmark success reflects genuine reasoning ability or sensitivity to superficial cues in the original data. In this sense, perturbation-based evaluation overlaps with adversarial evaluation, where inputs are intentionally altered to induce failure, but it is often used more broadly as a diagnostic tool for studying consistency and generalization under controlled changes \cite{wang2022advglue,Li2024PertEvalUR,Alzahrani2024WhenBA}.

For arithmetic and mathematical reasoning, perturbation-based evaluation is especially useful because relatively small changes can preserve the underlying problem while still exposing brittle behavior. SVAMP was designed to reveal shallow heuristics in arithmetic word-problem solving through carefully constructed variations of existing problems \cite{Patel2021AreNM}, while GSM-Plus studies the robustness of large language models on perturbed math word problems more systematically \cite{li2024gsmplus}. GSM-Symbolic similarly questions whether strong GSM8K performance reflects stable mathematical reasoning by evaluating models on symbolic variants of benchmark problems \cite{Mirzadeh2024GSMSymbolicUT}. More recently, MATH-Perturb extends this line of evaluation to harder mathematical settings by measuring performance under controlled problem modifications \cite{huang2025mathperturb}.

Numerical perturbation has also been studied directly as a test of whether models reason over arithmetic structure or reproduce learned numerical patterns. Yang et al. propose a template-based method for producing large-scale numerical variants of math word problems and show that LLMs remain vulnerable to numerical variation, often failing in arithmetic operations or producing invalid reasoning steps \cite{yang-etal-2025-evaluating}. This line of work motivates the need for evaluations that go beyond original benchmark instances and test whether models remain correct when the numbers in a problem change.

Our work is most closely aligned with these perturbation-based evaluations, but focuses specifically on schema-preserving numeric remapping for arithmetic word problems. Rather than generating variants only from manually specified templates, our pipeline operates on individual benchmark instances: it extracts a symbolic representation, generates constrained numeric remappings, recomputes the attacked gold answer, realizes the transformed question through a structured surface edit plan, and retains only high-confidence attacks after validation and post-hoc auditing. This allows us to test whether models remain correct on controlled, recomputed variants of problems they or the source model originally solved.

\subsection{Structured reasoning and procedural methods}
A large body of work aims to improve reasoning in language models by adding structure to intermediate problem solving rather than relying only on a direct final answer. Chain-of-thought prompting encourages models to produce step-by-step natural-language rationales and has been shown to improve performance on arithmetic and symbolic reasoning tasks \cite{wei2022cot}. Subsequent work showed that similar behavior can be elicited even in zero-shot settings with simple prompting cues \cite{kojima2023zeroshot}, while self-consistency improves reliability by sampling multiple reasoning traces and selecting the most consistent final answer \cite{Wang2022SelfConsistencyIC}. Other methods introduce stronger structure through decomposition or search. Least-to-Most prompting breaks difficult problems into simpler subproblems solved sequentially \cite{zhou2023leasttomost}, while Tree-of-Thoughts and Graph-of-Thoughts explore multiple candidate intermediate reasoning paths rather than committing to a single linear trace \cite{yao2023treethoughts,besta2024graphthoughts}.

A related direction improves intermediate reliability by delegating parts of reasoning to executable tools or external feedback. Program-aided language models and Program-of-Thoughts express parts of a solution as code that can be executed and checked, reducing arithmetic and symbolic errors \cite{gao2023pal,chen2023programthoughts}. Tool-augmented frameworks such as ReAct interleave reasoning with actions that retrieve information or query external resources, allowing intermediate claims to be grounded against observations \cite{yao2023react}. Other approaches focus on iterative improvement: Self-Refine and Reflexion use critique, feedback, or memory from earlier attempts to revise an initial solution and improve performance over repeated trials \cite{madaan2023selfrefine,shinn2023reflexion}.

Despite these advances, most prior methods still represent intermediate reasoning as free-form text, loosely structured search traces, or task-specific executable artifacts. This makes it difficult to verify whether intermediate steps preserve problem structure or to localize where reasoning fails under controlled perturbations. These limitations are especially relevant for arithmetic word problems, where schema-preserving changes can alter surface form without changing the underlying task.




\section{Attack Taxonomy}\label{sec:attack}
We study schema-preserving attacks: transformations that systematically alter an arithmetic word problem while preserving enough of the underlying task structure for the transformed example to remain coherent, solvable, and correctly labeled. The goal of these attacks is not simply to make problems harder, but to test whether a model's success reflects robust reasoning over the problem schema or brittle reliance on familiar benchmark patterns.

This section situates numeric remapping within a broader taxonomy of possible schema-preserving transformations. The taxonomy is useful because different attack families stress different aspects of arithmetic reasoning, including sensitivity to surface wording, numeric values, irrelevant information, relation changes, and target-variable changes. However, the experimental scope of this paper is limited to \textbf{numeric remapping}. The remaining families are included to define the broader transformation space and to clarify how numeric remapping relates to other possible generalization attacks.

Table~\ref{tab:attack_taxonomy} summarizes the attack families, the aspect of the example modified by each family, the intended invariant, and the corresponding generation and validation requirements.

\begin{table}[H]
\begin{adjustwidth}{-1cm}{-1cm}
\centering
\captionsetup{width=1.0\textwidth}
\caption{Taxonomy of schema-preserving attack families for arithmetic word problems. Each family changes a different aspect of the original problem while aiming to preserve a corresponding notion of task validity. The present paper implements and evaluates numeric remapping; the remaining families define the broader transformation space for future extensions.}
\label{tab:attack_taxonomy}
\footnotesize
\renewcommand{\arraystretch}{1.15}
\setlength{\tabcolsep}{3pt}
\begin{tabular}{
>{\raggedright\arraybackslash}p{1.8cm}
>{\raggedright\arraybackslash}p{1.9cm}
>{\raggedright\arraybackslash}p{2.2cm}
>{\raggedright\arraybackslash}p{2.8cm}
>{\raggedright\arraybackslash}p{3.0cm}
>{\raggedright\arraybackslash}p{1.2cm}
>{\raggedright\arraybackslash}p{1.3cm}
}
\toprule
\makecell[l]{\textbf{Family}} &
\makecell[l]{\textbf{Trans-}\\\textbf{formation}} &
\makecell[l]{\textbf{Invariant}} &
\makecell[l]{\textbf{Generation}} &
\makecell[l]{\textbf{Validation}} &
\makecell[l]{\textbf{Diffi-}\\\textbf{culty}} &
\makecell[l]{\textbf{Useful-}\\\textbf{ness}} \\
\midrule

Number Remapping
& Change numeric constants
& Keep reasoning program fixed
& Generate new values under local constraints and rewrite the text.
& Validate by substituting into the original symbolic computation and recomputing the gold answer.
& Low
& High \\

Lexical paraphrasing 
& Change surface wording 
& Preserve original quantities, relations, target answer 
& Paraphrase the text while preserving entities, quantities, relations, and gold answer
& Validate through semantic equivalence and answer preservation with re-extracted symbolic program of paraphrased question
& Low / Medium
& Medium \\

Unit Conversion
& Change quantity representation
& Keep semantics fixed
& Rewrite quantities into equivalent units or forms.
& Validate by normalizing to a canonical representation before recomputing the answer.
& Medium
& High \\

Distractor Insertion
& Add irrelevant facts
& Keep relevant computation fixed
& Insert plausible but irrelevant information into the problem text.
& Validate by confirming distractors do not enter the symbolic solution program.
& Medium
& High \\

Relation Substitution
& Change quantitative relations
& Preserve overall schema, with program updated accordingly
& Replace relations such as ``half'' with ``third'' or ``twice'' with ``triple,'' then propagate the change through the symbolic program.
& Validate against the transformed program.
& Medium / High
& High \\

Question Inversion
& Given the answer, try to solve for original question
& Keep core schema and internal consistency
& Reformulate the problem so a derived quantity becomes the target of inference.
& Validate by solving the transformed symbolic program for the new target.
& High
& High \\

Question Merging
& Combine compatible problems
& Preserve constituent reasoning patterns
& Merge two compatible problems or subproblems into one example.
& Validate by checking that each required subcomputation remains recoverable.
& High
& Medium / High \\

\bottomrule
\end{tabular}
\end{adjustwidth}
\end{table}

\textbf{Number remapping} changes the concrete numeric values in a problem while preserving the original reasoning program. This attack tests whether models have learned the underlying arithmetic schema or are instead sensitive to the particular quantities seen in the benchmark instance. Because the input values change, the attacked gold answer must be recomputed from the transformed symbolic representation. This is the attack family implemented and evaluated in the present paper.

\textbf{Lexical paraphrasing} rewrites the surface wording of a problem while preserving the original quantities, entities, relations, question target, and gold answer. Unlike numeric remapping, this attack does not change the symbolic computation or require a new answer; instead, it tests whether models remain robust to semantically equivalent rewordings of the same reasoning task. We include lexical paraphrasing in the taxonomy as a natural answer-preserving attack family, but do not report lexical-paraphrase experiments in this paper.

\textbf{Unit conversion} changes how quantities are expressed while preserving their underlying values after normalization. For example, a problem may replace hours with minutes or dollars with cents while leaving the mathematical relationship unchanged under the appropriate conversion. This attack tests whether models can reason over equivalent quantitative representations rather than relying on familiar surface forms.

\textbf{Distractor insertion} adds irrelevant but plausible information to the problem while preserving the quantities and relations needed to solve the original task. The gold answer remains unchanged because the added information should not enter the solution program. This attack tests whether models can identify the relevant reasoning path in the presence of semantically coherent but nonessential details.

\textbf{Relation substitution} changes a local quantitative relation while preserving the broader problem template. For example, a relation such as ``half as many'' may be replaced with ``a third as many,'' or an additive comparison may be replaced with a different compatible relation. Unlike lexical paraphrasing, this family changes the underlying computation, so the attacked gold answer must be recomputed from the transformed relation.

\textbf{Question inversion} changes which quantity is treated as the target of inference while preserving the underlying arithmetic relationships. Instead of asking for the original final quantity, the transformed problem provides enough information to solve for a different variable in the same symbolic schema. This attack tests whether models can reason flexibly over the structure of the problem rather than only following the original forward computation.

\textbf{Question merging} combines two or more compatible problem structures into a single attacked instance while preserving the recoverability of the intended subcomputations. This family tests whether models can maintain and compose multiple arithmetic schemas in a coherent word problem. Because merged examples require additional compatibility checks, this family is best viewed as a future extension of the attack-generation framework.

In the current paper, we instantiate this taxonomy through numeric remapping only. Numeric remapping is a useful first case because it preserves the original reasoning program while changing the concrete values that instantiate that program. This makes the transformation structured enough to validate automatically: the pipeline can check that the original symbolic program recovers the original gold answer, substitute new values into the same program, and recompute the attacked gold answer. The remainder of the paper therefore focuses on the generation, validation, and evaluation of numeric-remapping attacks.

\section{Methodology}\label{sec:methodology}
\subsection{Problem Setting}
We study numeric-remapping generalization attacks on arithmetic word problems. Let \(x\) denote an original problem instance with gold answer \(y\). A numeric-remapping attack generator \(A\) transforms \(x\) into an attacked instance \(x' = A(x)\), together with an attacked gold answer \(y'\). Unlike arbitrary perturbations, the goal is not simply to alter the input, but to generate a transformed example that remains a valid instance of the same underlying arithmetic task.

A numeric-remapping attack changes the concrete quantities in a problem while preserving the original reasoning program. For example, if the original problem requires applying a fixed arithmetic relationship among quantities, the attacked problem should instantiate the same relationship using different values. Because the input quantities change, the attacked gold answer must be recomputed rather than copied from the original example.

We say that an attacked example \((x', y')\) is valid if it satisfies three conditions. First, it must preserve the original problem schema, so that the attacked example corresponds to the same underlying reasoning pattern as the source problem. Second, it must remain internally consistent and solvable as a standalone word problem: the quantities, relations, and final question must remain coherent after remapping. Third, it must admit a verified attacked gold answer \(y'\), computed from the transformed symbolic representation rather than guessed heuristically.

This formulation distinguishes numeric remapping from unrestricted adversarial perturbation. We do not treat any performance drop as evidence of failure unless the transformed problem remains meaningful, valid, and correctly labeled. The methodology below therefore focuses on generating numeric-remapping attacks with stage-wise validation and recomputed gold answers.

\subsection{Automatic Numeric-Remapping Attack Generation Pipeline}
\begin{figure}[htbp]
    \centering
    \includegraphics[width=\linewidth]{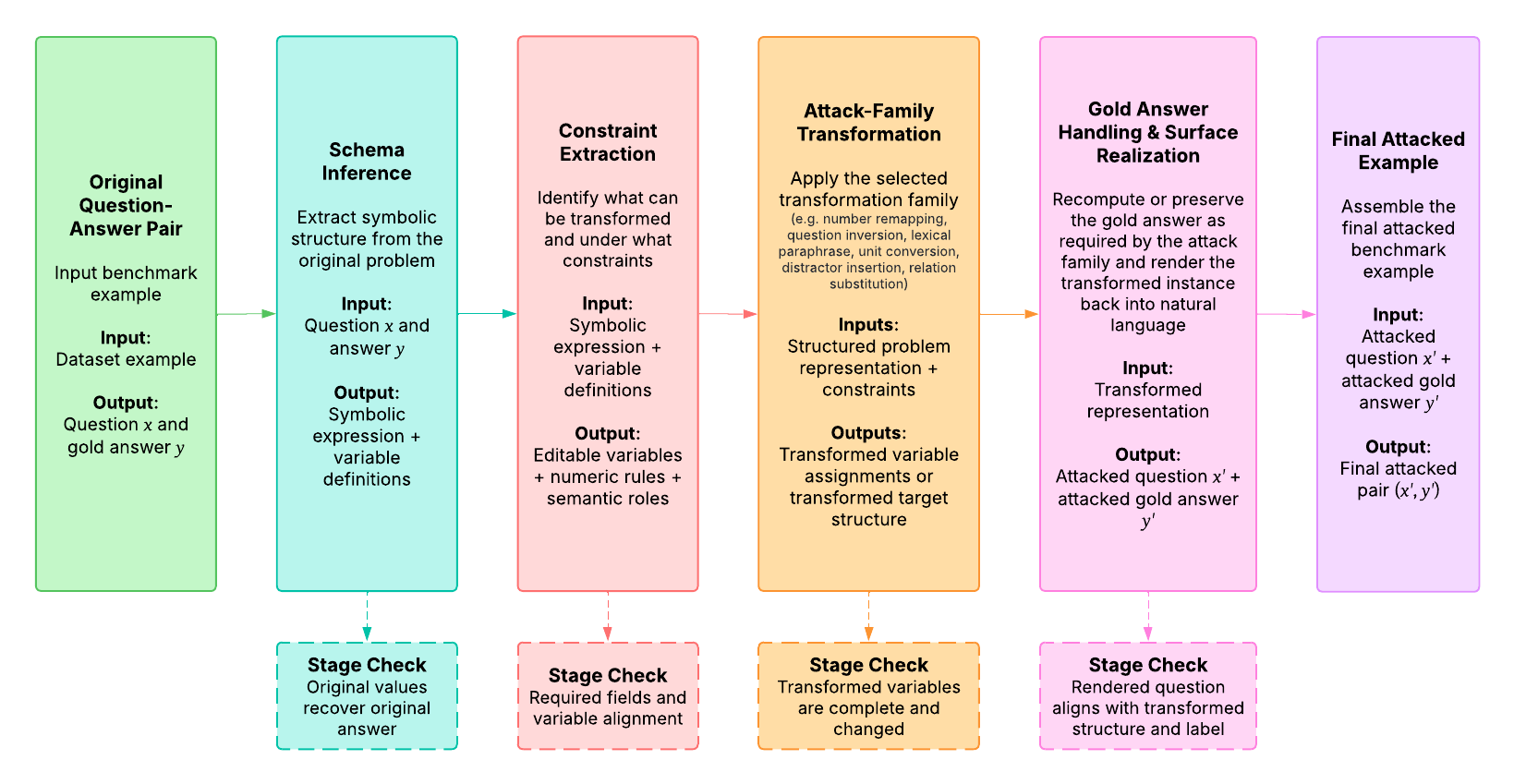}
    \caption{Overview of the numeric-remapping attack-generation pipeline. Starting from an original question-answer pair, the pipeline infers a symbolic problem representation, extracts transformation constraints, generates new numeric assignments, recomputes the attacked gold answer, and realizes the attacked question through an LLM-generated surface edit plan that is applied deterministically. Intermediate artifacts, including the final rendered question, are checked incrementally as they are produced and passed to the next stage.}
    \label{fig:attack_pipeline}
\end{figure}

Our attack-generation pipeline transforms an original benchmark example into a numeric-remapped attacked example through a sequence of structured stages. Starting from the original question and answer, the pipeline first infers a symbolic representation of the problem schema, extracts constraints on editable quantities, proposes new values under those constraints, recomputes the attacked gold answer, generates a structured surface edit plan, and deterministically applies that plan to produce the attacked question. Figure~\ref{fig:attack_pipeline} summarizes this process.

A key design choice is that correctness is enforced incrementally rather than through a single final validation step. Each stage emits a structured artifact that is checked before being passed forward. In this way, failures in schema extraction, constraint construction, numeric remapping, or surface realization can be detected early, preventing invalid intermediate products from propagating into the final attacked example.

\begin{algorithm}[htbp]
\caption{Numeric-Remapping Attack Generation}
\label{alg:numeric_remapping}
\begin{algorithmic}[1]
\Require Source model \(m\), generation model \(\ell\), dataset \(D\), schema retry budget \(B_s\), constraint retry budget \(B_c\), remapping retry budget \(B_r\), surface retry budget \(B_e\)
\Ensure Valid numeric-remapping attack set \(D'\)

\State \(D' \gets \emptyset\)

\ForAll{\((x,y) \in D\)}
    \State \(\hat{y} \gets m(x)\)
    \If{\(\hat{y} \neq y\)}
        \State \textbf{continue}
    \EndIf

    \State \(S \gets \Call{InferSchema}{\ell, x, y, B_s}\)
    \If{\(\neg \Call{ValidSchema}{S, x, y}\)}
        \State \textbf{continue}
    \EndIf

    \State \(C \gets \Call{ExtractConstraints}{\ell, x, y, S, B_c}\)
    \If{\(\neg \Call{ValidConstraints}{C, S}\)}
        \State \textbf{continue}
    \EndIf

    \State \(R \gets \Call{GenerateRemap}{\ell, x, y, S, C, B_r}\)
    \If{\(\neg \Call{ValidRemap}{R, S, C}\)}
        \State \textbf{continue}
    \EndIf

    \State \(y' \gets \Call{EvaluateSymbolicProgram}{S, R}\)

    \State \(E \gets \Call{GenerateSurfaceEditPlan}{\ell, x, S, C, R, B_e}\)
    \If{\(\neg \Call{ValidEditPlan}{E, x, R}\)}
        \State \textbf{continue}
    \EndIf

    \State \(x' \gets \Call{ApplySurfaceEdits}{x, E}\)

    \If{\(\Call{ValidRenderedAttack}{x', y', S, C, R, E}\)}
        \State \(D' \gets D' \cup \{(x, y, x', y', S, C, R, E)\}\)
    \EndIf
\EndFor

\State \Return \(D'\)

\end{algorithmic}
\end{algorithm}

Algorithm~\ref{alg:numeric_remapping} summarizes the numeric-remapping attack-generation procedure. The main text describes each stage at a high level; additional details about the validation checks used at each stage are provided in Appendix~\ref{app:stage_checks}.

\subsubsection{Schema Inference}
The first stage constructs a structured representation of the original problem that captures the quantities, relations, and computation needed to derive the correct answer. In our setting, this representation consists of a symbolic expression together with a dictionary of semantically meaningful input variables. Each variable corresponds to a quantity that appears in the original problem and is associated with metadata such as its meaning and original value.

The purpose of this stage is not merely to reproduce the final answer, but to recover the arithmetic structure of the original example in a form that can be manipulated downstream. This symbolic representation serves as the anchor for numeric remapping because it allows later stages to modify selected quantities while tracking how those changes propagate through the solution structure.

To ensure that the extracted representation is trustworthy enough to support attack generation, it is verified against the original example. In particular, substituting the original variable values into the symbolic expression must reproduce the original gold answer. Only examples whose structured representation passes this verification step are retained for downstream remapping.

\subsubsection{Constraint Extraction}
Once a verified symbolic representation has been obtained, the pipeline augments each input variable with metadata needed for controlled numeric transformation. For each quantity, this metadata may include its source text, numeric type, replaceability, semantic role in the problem, and any local numeric or semantic constraints that must be respected during remapping.

In practice, these constraints may specify whether a variable must remain fixed, whether it should remain integer-valued, whether it must remain positive, or whether it plays a particular semantic role such as a base quantity, offset, count, price, rate, or multiplicative factor. The purpose of this stage is to distinguish which components of the problem may be transformed, which must remain fixed, and what conditions the transformation must satisfy in order for the attacked problem to remain valid.

This stage produces a structured constraint specification rather than a free-form description. The resulting artifact provides the remapping stage with an explicit account of the represented quantities and the conditions under which they may be changed. As in the previous stage, this artifact is checked before being passed forward, so missing fields, malformed outputs, or incompatible variable specifications are rejected early.

\subsubsection{Numeric Remapping}
The numeric-remapping stage proposes new values for editable variables while preserving the original reasoning program. The transformed values are chosen subject to the structured constraints extracted in the previous step. As a result, the attacked instance continues to instantiate the same underlying arithmetic structure, but with different concrete quantities.

The output of this stage is a remapping specification that assigns new values to editable variables and, when needed, provides preferred replacement text for surface realization. This replacement text describes how the variable should be expressed when it appears directly in the problem. However, because the same variable may also be expressed through additional co-referential mentions, the downstream surface-realization stage is responsible for identifying all text spans that must be updated. For example, a variable whose original source text is ``half'' may be represented numerically as \(0.5\) or through a divisor value of \(2\), but its replacement must preserve the relational meaning in natural language. A valid remapping should therefore rewrite the relation coherently rather than simply inserting the raw numeric value into the text.

Before a remapping is accepted, the pipeline checks that the transformed assignment satisfies the expected structure. Editable variables must receive valid replacement values, fixed variables must remain unchanged, and the remapped values must satisfy the extracted numeric and semantic constraints. The transformed symbolic expression is then evaluated under the new assignments to compute the attacked gold answer.

\subsubsection{Gold Answer Recomposition and Surface Realization}
After a valid remapping has been produced, the attacked gold answer is recomputed from the symbolic representation. Let \(f\) denote the symbolic expression extracted from the original problem and let \(r\) denote the new variable assignment produced by the remapping stage. The attacked answer is computed by evaluating \(f(r)\). Thus, the attacked label is not produced by the language model and is not copied from the original example; it is derived from the verified arithmetic structure.

The pipeline then renders the attacked problem back into natural language. A central challenge in this stage is that a single symbolic variable may be expressed by more than one surface mention in the original question. For example, a chair-capacity variable may appear once as ``a capacity of two people each'' and later again as ``the rest each had two people.'' Replacing only the primary source span would leave a stale co-referential mention in the rendered question, producing an attacked example whose text no longer matches the recomputed answer.

To address this issue, surface realization is performed in two steps. First, the generation model is prompted to produce a structured edit plan rather than a free-form rewritten question. Given the original question, symbolic representation, constraints, and remap, the model identifies exact text spans in the original question that must be changed and proposes replacement text for each span. The edit plan may include both direct mentions of remapped quantities and co-referential mentions that express the same underlying variable. The model is instructed not to rewrite the full problem, but instead to preserve all unchanged text exactly and return only structured edit actions.

Second, the edit plan is applied deterministically by code. Each proposed source span must appear in the original question, edit spans must be non-overlapping, and replacements are applied to the original text rather than to an already modified intermediate string. This design uses the language model for the semantic task of identifying all relevant surface mentions, while keeping the actual text transformation controlled and reproducible.

The final rendered question is then checked against the remapping and recomputed answer. These checks verify that required remapped quantities are reflected in the rendered text, stale source spans are not left behind, malformed surface patterns are not introduced, and the recomputed symbolic answer agrees with the stored attacked label. These checks are practical filtering criteria rather than a proof of perfect semantic equivalence, but they substantially reduce invalid attacks and ensure that retained examples satisfy the numeric-remapping definition used in this paper.

\subsection{Worked Example}
To make the pipeline concrete, we illustrate numeric remapping using a representative GSM8K example.

\subsubsection{Source Example and Schema Inference}
\begin{quote}
\textbf{Question:} Natalia sold clips to 48 of her friends in April, and then she sold half as many clips in May. How many clips did Natalia sell altogether in April and May?

\textbf{Gold answer:} 72
\end{quote}

From the original question-answer pair, the pipeline extracts a symbolic representation of the required computation:
\[
\texttt{final} = \texttt{april\_clips} + \frac{\texttt{april\_clips}}{\texttt{divisor}},
\]
with variables
\[
\texttt{april\_clips} = 48, \qquad \texttt{divisor} = 2.
\]

Here, \texttt{april\_clips} represents the number of clips sold in April, and \texttt{divisor} captures the relational phrase ``half as many.'' Substituting the original values yields
\[
48 + 48/2 = 72,
\]
which matches the original gold answer. The extracted symbolic artifact is therefore retained for numeric remapping.

\subsubsection{Constraint Extraction}
The pipeline next identifies which quantities can be modified and under what conditions. In this example, \texttt{april\_clips} is treated as a replaceable integer count, while \texttt{divisor} is treated as a replaceable relational quantity governing how May sales are derived from April sales. The constraint artifact records that the replacement values must preserve the semantics of the story and remain compatible with the symbolic expression.

\subsubsection{Numeric Remapping and Recomputed Answer}
The remapping stage proposes new values
\[
\texttt{april\_clips} = 30, \qquad \texttt{divisor} = 5.
\]

The attacked gold answer is then recomputed from the same symbolic program:
\[
30 + 30/5 = 36.
\]

The surface-realization stage expresses these remapped values through a structured edit plan. In this example, the edit plan replaces the April quantity with ``30 of her friends'' and replaces the relational phrase ``half as many'' with ``a fifth as many.'' The second edit is important because the symbolic variable \texttt{divisor} does not correspond to a standalone quantity in the story; it is expressed through a natural-language relation. The rendered question must therefore update the relation itself, not simply insert the numeric value \(5\).

The resulting attacked example is:

\begin{quote}
\textbf{Question:} Natalia sold clips to 30 of her friends in April, and then she sold a fifth as many clips in May. How many clips did Natalia sell altogether in April and May?

\textbf{Gold answer:} 36
\end{quote}

This attacked example is valid because the surface text, symbolic representation, and recomputed answer all express the same transformed problem: May sales are one fifth of April sales, so the total is \(30 + 30/5 = 36\). By contrast, a question such as ``Natalia sold clips to 30 of her friends in April, and then she sold 5 clips in May'' would be invalid for this remapping, because it changes the meaning from a proportional relation to an absolute quantity.

\subsection{Scope of the Current Implementation}
The broader taxonomy in Section~\ref{sec:attack} describes several possible schema-preserving transformations for arithmetic word problems. In this paper, however, the implemented pipeline and empirical evaluation focus on numeric remapping only. This scope allows us to study a controlled answer-changing transformation for which the attacked label can be recomputed directly from the preserved symbolic program.

Other attack families would require different transformation and validation procedures. For example, some transformations may preserve the original answer, while others may require changing the target quantity or modifying the symbolic program itself. We leave these extensions to future work and restrict the experiments in this paper to numeric-remapping attacks.

\section{Experiments}\label{sec:experiments}
\subsection{Computational Infrastructure and Runtime Comparison}
\label{sec:runtime_comparison}

Our experimental pipeline is computationally intensive. Beyond standard benchmark evaluation, numeric-remapping attack generation requires multiple structured language-model calls per example, stage-wise validation, and bounded retry/backtracking when intermediate outputs fail. In practice, this means that a single attacked benchmark may require many thousands of large-model inference calls before a final validated dataset is produced.

To support these workloads we mainly use the Supermicro ARS-111GL-NHR system, abbreviated as \textbf{GH200 Supermicro}. Then, we compare the time performances of \textbf{GH200 Supermicro} with the HP ZGX Nano G1n AI Station, abbreviated as \textbf{ZGX Nano}. 

\begin{table}[htbp]
\centering
\small
\caption{Hardware comparison for the machines used in this work. For comparability, we report both system RAM and accelerator-accessible memory. The ZGX Nano uses GB10 unified memory, whereas the GH200 Supermicro exposes separate system and accelerator memory regions.}
\label{tab:hardware_comparison}
\setlength{\tabcolsep}{5pt}
\renewcommand{\arraystretch}{1.5}
\begin{tabular}{>{\raggedright\arraybackslash}p{2.8cm}
                >{\raggedright\arraybackslash}p{2.2cm}
                c
                >{\raggedright\arraybackslash}p{2.2cm}
                >{\raggedright\arraybackslash}p{2.3cm}
                >{\raggedright\arraybackslash}p{2.6cm}}
\toprule
\textbf{Machine} & \textbf{CPU} & \textbf{Cores} & \textbf{System RAM} & \textbf{GPU} & \textbf{Accelerator Memory} \\
\midrule
ZGX Nano & Cortex-X925/A725 & 20 & 119\,GiB & NVIDIA GB10 & 128\,GB unified \\
GH200 Supermicro & Neoverse-V2 & 72 & 573\,GiB & NVIDIA GH200 480GB & 97\,GB accelerator-attached \\
\bottomrule
\end{tabular}
\end{table}

Table~\ref{tab:hardware_comparison} summarizes the relevant hardware characteristics of the two systems. Relative to the ZGX Nano, the GH200 Supermicro provides substantially greater CPU and system-memory capacity, increasing from 20 CPU cores and 119\,GiB of RAM to 72 CPU cores and 573\,GiB of RAM. The two systems differ in memory architecture as well: the ZGX Nano uses GB10 unified memory, while the GH200 Supermicro exposes a separate accelerator-attached memory region. This additional capacity is directly relevant to our workload, which combines repeated benchmark evaluation with iterative multi-stage attack generation.

\begin{table}[htbp]
\centering
\caption{Observed runtime comparison for MAWPS numeric-remapping attack generation. Both runs use the same source set of 1{,}700 attempted examples, the same GPT-OSS (120B) generation model, and the same pipeline settings. Speedup is measured relative to the ZGX Nano.}
\label{tab:runtime_comparison}
\setlength{\tabcolsep}{4pt}
\renewcommand{\arraystretch}{1.2}
\begin{tabular}{>{\raggedright\arraybackslash}p{3.2cm}
                >{\raggedright\arraybackslash}p{2.3cm}
                >{\raggedright\arraybackslash}p{1.9cm}
                >{\raggedright\arraybackslash}p{3.4cm}
                >{\raggedright\arraybackslash}p{1.8cm}}
\toprule
\textbf{Machine} & \textbf{Examples Attempted} & \textbf{Total Time} & \textbf{Avg. Time/Example (s)} & \textbf{Speedup} \\
\midrule
ZGX Nano & 1{,}700 & 29h 24m 11s & 62.27 & 1.00$\times$ \\
GH200 Supermicro & 1{,}700 & 8h 11m 35s & 17.35 & 3.59$\times$ \\
\bottomrule
\end{tabular}
\end{table}

We directly compare runtime for the same MAWPS numeric-remapping generation workload on the two systems. Both runs use the same source set of 1{,}700 attempted examples, the same generation model, and the same pipeline settings. Table~\ref{tab:runtime_comparison} reports total runtime, average time per example, and speedup relative to the ZGX Nano. On this workload, the GH200 Supermicro achieves a speedup of \(3.59\times\), reducing the average processing time from over a minute to about 17 seconds per example.

To contextualize how this speedup affects full attack-set construction across datasets, we compare dataset-level generation times for numeric-remapping attack construction on the ZGX Nano and GH200 Supermicro. For the ZGX Nano, runtimes are observed for all three datasets: GSM8K, MAWPS, and MultiArith. For the GH200 Supermicro, the MAWPS runtime is directly measured.

\begin{figure}[!htbp]
    \centering
    \includegraphics[width=0.9\linewidth]{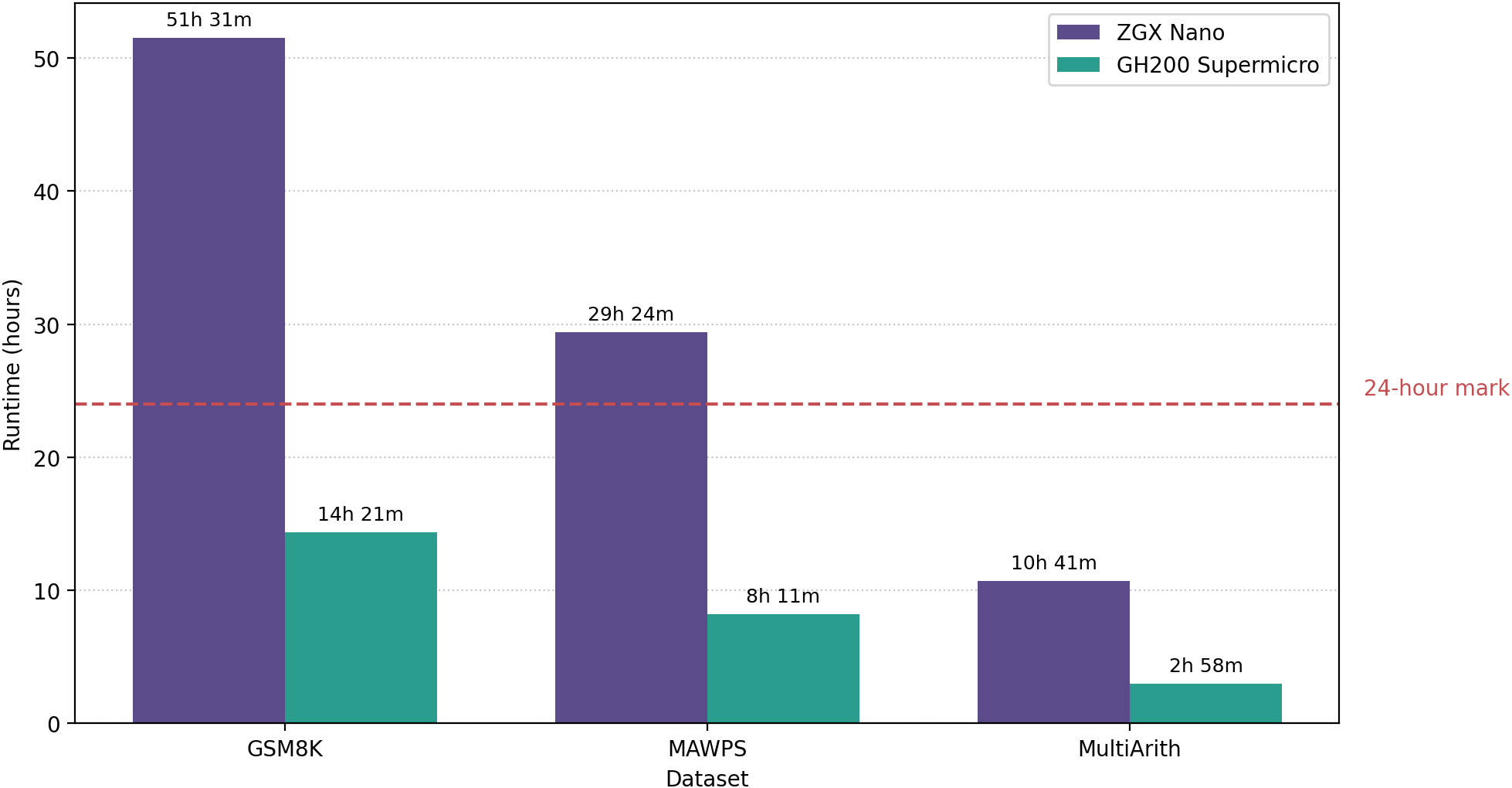}
    \caption{Numeric-remapping attack-generation runtime by dataset on the ZGX Nano and GH200 Supermicro. Attacks are generated with GPT-OSS (120B). The dashed horizontal line marks 24 hours, highlighting that the ZGX Nano requires more than a day for the larger GSM8K and MAWPS attack-generation workloads, while the GH200 Supermicro reduces those workloads to a handful of hours.}
    \label{fig:runtime_comparison_by_dataset}
\end{figure}

\begin{figure}[!htbp]
    \centering
    \includegraphics[width=0.8\linewidth]{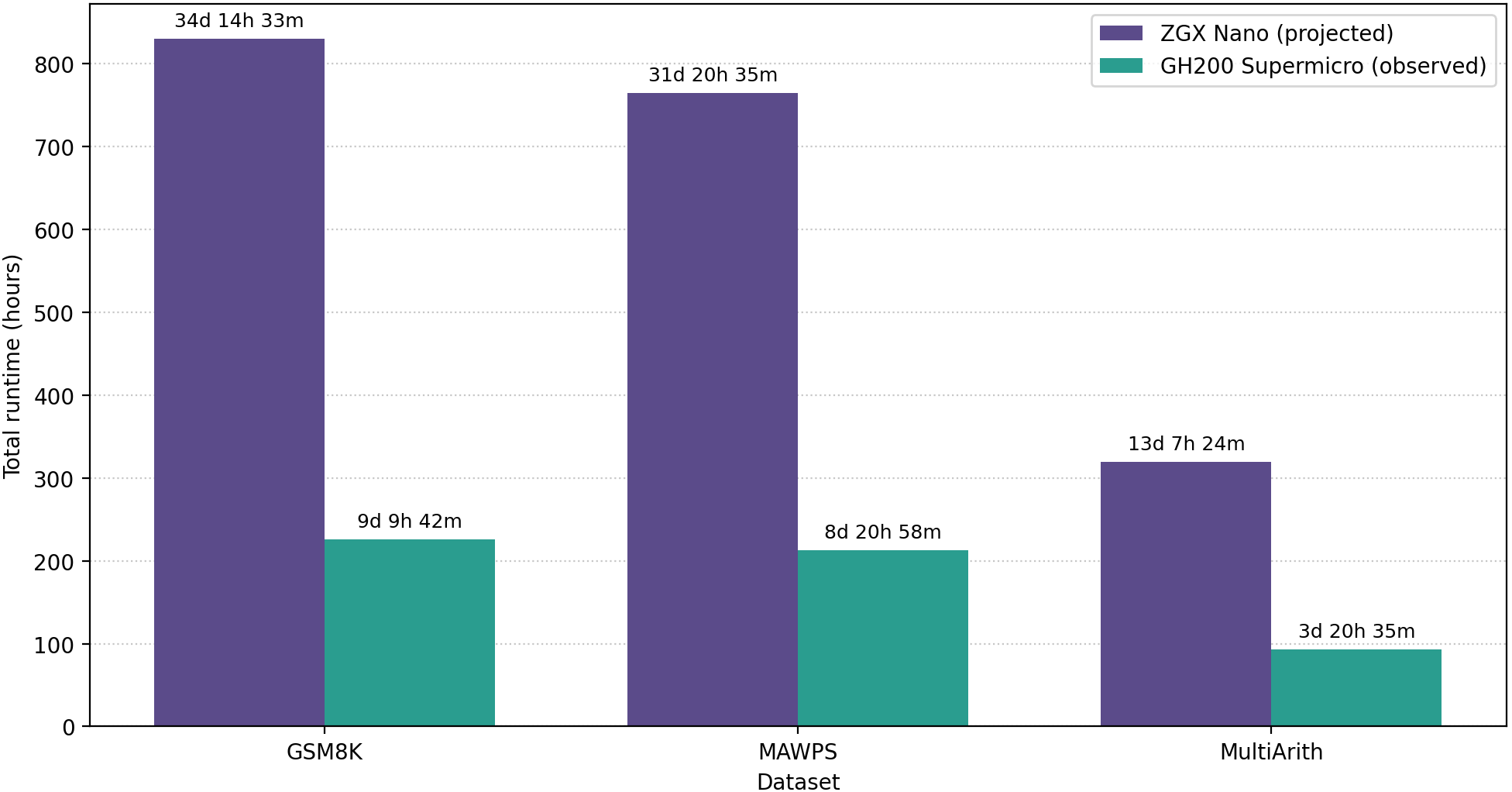}
    \caption{Baseline evaluation runtime for DeepSeek-R1 (70B).}
    \label{fig:deepseek_runtime_comparison}
\end{figure}

\begin{figure}[!htbp]
    \centering
    \includegraphics[width=0.8\linewidth]{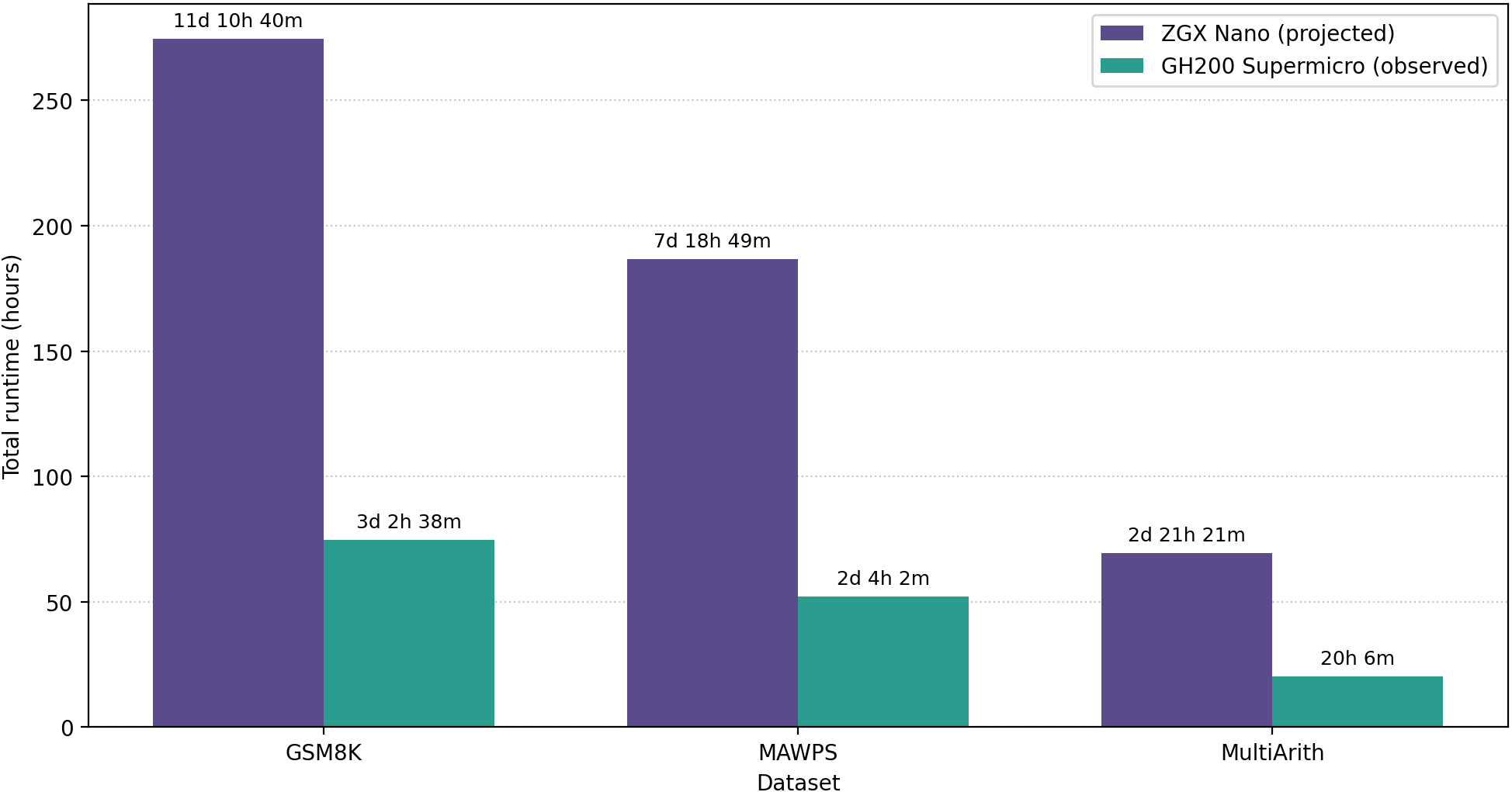}
    \caption{Baseline evaluation runtime for Gemma4 (31B).}
    \label{fig:gemma_runtime_comparison}
\end{figure}

\begin{figure}[!htbp]
    \centering
    \includegraphics[width=0.8\linewidth]{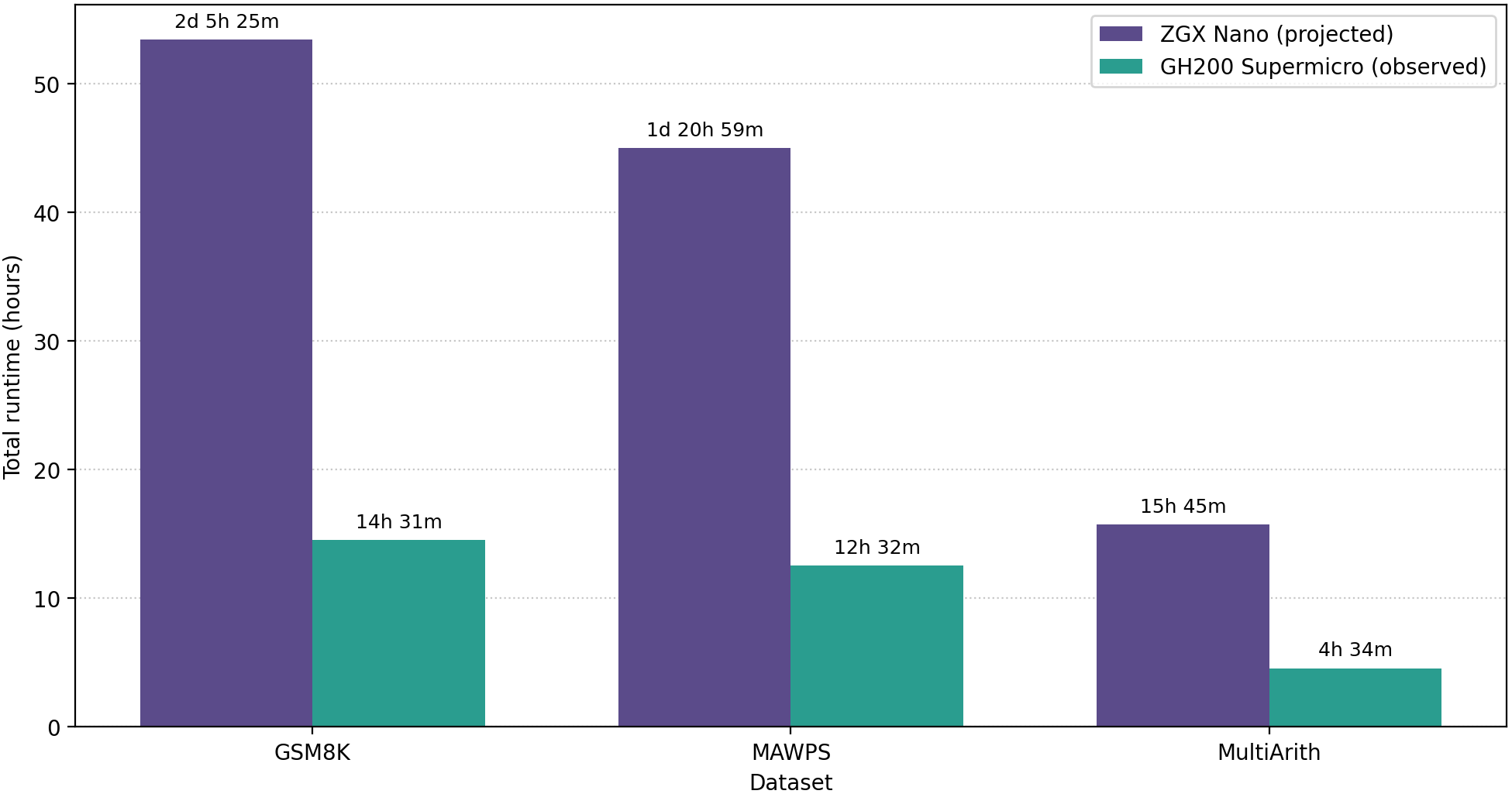}
    \caption{Baseline evaluation runtime for GPT-OSS (120B).}
    \label{fig:gptoss_runtime_comparison}
\end{figure}

Across datasets, this comparison illustrates that the GH200 Supermicro substantially reduces the turnaround time required to construct numeric-remapping attack sets. For GSM8K, the ZGX Nano requires approximately 51 hours of generation time. For MAWPS, the observed runtime decreases from 29h 24m on the ZGX Nano to 8h 11m on the GH200 Supermicro. For MultiArith, the observed ZGX Nano runtime is approximately 10h 41m. This difference makes iterative debugging, validation, regeneration, and re-evaluation substantially more practical.

In addition to attack-generation runtime, we also compare the cost of baseline model evaluation across machines. The baseline evaluations in this study were run on the GH200 Supermicro. Figures~\ref{fig:deepseek_runtime_comparison}--\ref{fig:gptoss_runtime_comparison} show the resulting runtime comparisons for DeepSeek-R1, Gemma4, and GPT-OSS. These figures illustrate that model evaluation itself can become a major computational bottleneck, especially for slower reasoning models such as DeepSeek-R1.

\subsection{Attack Set Construction}
\label{sec:attack_set_construction}
Attack sets are constructed using the staged numeric-remapping pipeline described in Section~\ref{sec:methodology}. We first run all evaluated models on each fixed dataset subset to obtain baseline results. Because GPT-OSS (120B) achieves the strongest baseline performance across the evaluated datasets, we use the subset of examples it answers correctly as the source set from which attacks are generated. Using a single strongest source model to define the source subset allows us to construct one attacked benchmark per dataset and evaluate all target models on the same transformed examples.

This yields \(1{,}885\) source examples for GSM8K, \(1{,}700\) for MAWPS, and \(593\) for MultiArith. These source examples are then passed through the numeric-remapping attack-generation pipeline. An example is first retained if it successfully passes all required generation stages: schema inference, constraint extraction, numeric remapping, surface edit-plan generation, deterministic surface realization, and rendered-question validation.

For schema inference, acceptance requires that the generated symbolic program execute successfully, that extracted variables align with the symbols appearing in the expression, and that substituting the original variable assignments recover the original gold answer. For constraint extraction, acceptance requires that all expected fields be present and that outputs satisfy the required type structure. For numeric remapping, acceptance requires that editable variables receive valid transformed values, fixed variables remain unchanged, and the remapped values satisfy extracted numeric and semantic constraints. For surface realization, acceptance requires that the edit plan refer to exact spans in the original question, that the spans be non-overlapping, and that the rendered question reflect the remapped values while preserving the intended problem schema.

After generation, we apply an additional post-hoc audit to construct the final high-confidence attack set used for model evaluation and reporting. This audit is intentionally conservative. It checks that the recomputed attacked answer agrees with the symbolic expression under the remapped values, that the rendered question is not unchanged from the source question when the answer changes, that required remapped quantities or replacement surfaces appear in the attacked question, and that obvious malformed surface patterns or stale source spans are not present. Attacks that fail these high-confidence criteria are excluded from the final evaluation set or marked for manual review rather than being counted as valid attacks.

This additional audit is important because surface realization can introduce subtle errors even when earlier structured artifacts are correct. For example, a remapped variable may be expressed through multiple co-referential surface mentions, and failing to update all of them can produce a question whose text no longer matches the recomputed answer. The final attacked benchmark for each dataset therefore consists only of examples that pass both the staged generation pipeline and the post-hoc high-confidence audit.

We report the resulting high-confidence attack-set sizes and retained attack rates in Section~\ref{sec:results}. These retained examples define the fixed attacked benchmark used for all target-model evaluations.

\subsection{Evaluation Metrics}
\label{sec:evaluation_metrics}
We evaluate model performance using both standard accuracy measures and numeric-remapping robustness metrics. For each model and dataset, we report baseline accuracy on the original benchmark subset. For attacked evaluation, we report performance on the fixed high-confidence numeric-remapped attack set generated from the GPT-OSS-correct source subset for that dataset.

Because the attacked benchmark is derived from examples that were originally answered correctly by the source model and that also pass the attack-generation and post-hoc audit pipeline, we report \emph{conditional attacked accuracy}: the fraction of retained high-confidence attacked examples that the model answers correctly. We also report \emph{conditional drop}, measured in percentage points, as the decrease from 100\% correctness on the retained source subset to the observed conditional attacked accuracy. Thus, a model with \(80\%\) conditional attacked accuracy has a conditional drop of \(20\) percentage points.

In addition to model accuracy, we report attack-set construction statistics that characterize the reliability of the numeric-remapping generation process itself. These include the number of source examples, the number of high-confidence retained attacked examples, and the resulting retention rate. Where available, we also report stage-wise pass/fail counts and failure statistics such as schema-extraction failures, remapping failures, surface-realization failures, and post-hoc audit exclusions. Together, these metrics capture both model robustness under numeric remapping and the reliability of the attack-construction pipeline.

Additional examples, validation criteria, structured output templates, and runtime calculation details are provided in Appendix~\ref{app:numeric_examples}--\ref{app:stage_checks}.

We now turn to the empirical results under this evaluation setup.

\section{Results}\label{sec:results}
\subsection{Baseline Results}
\label{sec:baseline_results}

We first evaluate each model on the original arithmetic word-problem benchmarks before applying numeric-remapping attacks. These baseline results establish the reference performance on the unmodified dataset subsets described in Section~\ref{sec:attack_set_construction}. Table~\ref{tab:baseline} reports the number of evaluated examples, the number answered correctly, accuracy, and average inference time per example.

\begin{table}[htbp]
\centering
\caption{Baseline accuracy on arithmetic reasoning benchmarks.
  $N$ is the number of evaluated examples; Avg.\ time is mean
  inference time per example.}
\label{tab:baseline}
\setlength{\tabcolsep}{6pt}
\begin{tabular}{llrrrr}
\toprule
\textbf{Model} & \textbf{Dataset} & $N$ & \textbf{Correct} & \textbf{Accuracy (\%)} & \textbf{Avg.\ time (s)} \\
\midrule
\multirow{3}{*}{DeepSeek-R1 (70B)}
  & GSM8K      & 2{,}000 & 1{,}659 & 82.95 & 406.25 \\
  & MAWPS      & 1{,}772 & 1{,}567 & 88.43 & 432.68 \\
  & MultiArith &   600   &   574   & 95.67 & 555.49 \\
\midrule
\multirow{3}{*}{Gemma4 (31B)}
  & GSM8K      & 2{,}000 & 1{,}627 & 81.35 & 134.35 \\
  & MAWPS      & 1{,}772 & 1{,}674 & 94.47 & 105.72 \\
  & MultiArith &   600   &   587   & 97.83 & 120.62 \\
\midrule
\multirow{3}{*}{GPT-OSS (120B)}
  & GSM8K      & 2{,}000 & 1{,}885 & 94.25 &  26.13 \\
  & MAWPS      & 1{,}772 & 1{,}700 & 95.94 &  25.46 \\
  & MultiArith &   600   &   593   & 98.83 &  27.39 \\
\bottomrule
\end{tabular}
\end{table}

Across the original benchmark subsets, the evaluated models achieve strong baseline performance, especially on MAWPS and MultiArith. GPT-OSS (120B) obtains the strongest baseline accuracy across the datasets and is therefore used as the source model for defining the baseline-correct subset from which numeric-remapping attacks are generated.

\subsection{Numeric-Remapping Attack Set Construction}
\label{sec:numeric_remapping_construction}
Because our goal is to test whether correct original reasoning survives schema-preserving numeric transformation, we generate attacks only from source questions that were answered correctly by GPT-OSS (120B) on the original benchmark. We further restrict evaluation to cases where the numeric-remapping pipeline produced a valid attacked example and where the resulting attacked question passed a post-hoc high-confidence audit. The resulting attacked set is therefore a conditional robustness benchmark: it measures whether models remain correct on valid numeric variants of problems that the source model originally solved.

For each dataset, we construct a single numeric-remapped benchmark by applying the attack-generation pipeline to the subset of examples answered correctly by GPT-OSS (120B). The same attacked set is then evaluated across all target models. This setup ensures that differences in attacked accuracy reflect model behavior on the same transformed examples rather than differences in attack generation across models.

\begin{table}[htbp]
\centering
\caption{Numeric-remapping attack-set construction after post-hoc high-confidence auditing. For each dataset, attacks are generated from the subset of examples answered correctly by the source model, GPT-OSS (120B). Retention rate is the fraction of source examples for which a high-confidence numeric-remapped version was retained after staged generation checks and post-hoc audit filtering.}
\label{tab:numeric_remapping_retention}
\setlength{\tabcolsep}{6pt}
\begin{tabular}{lrrr}
\toprule
\textbf{Dataset} & \textbf{Source Correct} & \makecell[c]{\textbf{High-Confidence}\\\textbf{Attacks}} & \textbf{Retention Rate (\%)} \\
\midrule
GSM8K      & 1{,}885 &   732 & 38.83 \\
MAWPS      & 1{,}700 & 1{,}597 & 93.94 \\
MultiArith &   593   &   570 & 96.12 \\
\bottomrule
\end{tabular}
\end{table}

Table~\ref{tab:numeric_remapping_retention} reports the number of source examples and retained high-confidence numeric-remapping attacks for each dataset. After post-hoc audit filtering, the pipeline retains \(732\) attacks from \(1{,}885\) GSM8K source examples, corresponding to a retention rate of \(38.83\%\). Retention is substantially higher for MAWPS and MultiArith, with \(1{,}597\) retained attacks from \(1{,}700\) MAWPS source examples and \(570\) retained attacks from \(593\) MultiArith source examples. These differences suggest that high-confidence numeric remapping is more difficult for GSM8K, likely because GSM8K examples are longer, more linguistically varied, and more likely to contain co-referential or relational surface forms that must be updated consistently.

\subsection{Numeric-Remapping Robustness Results}
\label{sec:numeric_remapping_results}
We next evaluate model performance on the high-confidence numeric-remapped attack sets. For each dataset, all models are evaluated on the same attacked examples generated from the GPT-OSS-correct source subset and retained after post-hoc auditing. We report \emph{conditional attacked accuracy}, defined as the fraction of retained attacked examples answered correctly under numeric remapping. We also report \emph{conditional drop}, measured in percentage points, as the decrease from \(100\%\) correctness on the retained source subset to the observed attacked accuracy.

\begin{table}[htbp]
\centering
\caption{Model performance on high-confidence numeric-remapping attacks. For each dataset, all models are evaluated on the same attacked set generated from examples originally answered correctly by GPT-OSS (120B) and retained after post-hoc high-confidence auditing. ``Cond. drop'' reports the conditional decrease from \(100\%\) original correctness on the retained source subset to the observed attacked accuracy, measured in percentage points (pp).}
\label{tab:numeric_remapping_results}
\setlength{\tabcolsep}{4.5pt}
\begin{tabular}{llrrrrr}
\toprule
\textbf{Model} & \textbf{Dataset} & \makecell[c]{\textbf{Valid}\\\textbf{attacks}} & \textbf{Correct} & \makecell[c]{\textbf{Accuracy}\\\textbf{(\%)}} & \makecell[c]{\textbf{Cond.}\\\textbf{drop (pp)}} & \makecell[c]{\textbf{Avg.}\\\textbf{time (s)}} \\
\midrule
\multirow{3}{*}{Gemma4 (31B)}
  & GSM8K      &   732 &   543 & 74.18 & 25.82 &  58.59 \\
  & MAWPS      & 1{,}597 & 1{,}569 & 98.25 &  1.75 &  32.55 \\
  & MultiArith &   570 &   561 & 98.42 &  1.58 &  24.03 \\
\midrule
\multirow{3}{*}{DeepSeek-R1 (70B)}
  & GSM8K      &   732 &   590 & 80.60 & 19.40 & 185.89 \\
  & MAWPS      & 1{,}597 & 1{,}543 & 96.62 & 3.38 & 146.06 \\
  & MultiArith &   570 &   560 & 98.25 &  1.75 & 111.46 \\
\midrule
\multirow{3}{*}{GPT-OSS (120B)}
  & GSM8K      &   732 &   643 & 87.84 & 12.16 &   8.07 \\
  & MAWPS      & 1{,}597 & 1{,}578 & 98.81 &  1.19 &   8.12 \\
  & MultiArith &   570 &   570 & 100.00 & 0.00 &   5.78 \\
\bottomrule
\end{tabular}
\end{table}

On GSM8K, the evaluated models exhibit the largest performance degradation under numeric remapping. Gemma4 falls to \(74.18\%\) conditional attacked accuracy, corresponding to a \(25.82\)-point conditional drop. DeepSeek-R1 performs better at \(80.60\%\), with a \(19.40\)-point drop. GPT-OSS remains the strongest completed GSM8K result at \(87.84\%\), but still loses \(12.16\) percentage points relative to the baseline-correct source subset. These results show that even after stricter high-confidence filtering, GSM8K numeric remapping exposes meaningful brittleness in arithmetic reasoning.

On MAWPS and MultiArith, the effect of numeric remapping is much smaller. For MAWPS, completed model results are above \(98\%\) conditional attacked accuracy, with GPT-OSS reaching \(98.81\%\), and Gemma4 reaching \(98.25\%\). For MultiArith, GPT-OSS answers all retained attacks correctly, while Gemma4 and DeepSeek-R1 remain between \(98.25\%\) and \(98.42\%\). These smaller drops suggest that robustness under numeric remapping depends strongly on dataset structure. Shorter and more regular arithmetic problems appear less vulnerable to this attack family, while GSM8K's longer and more varied language provides more opportunities for both attack-generation difficulty and model reasoning instability.

Overall, the numeric-remapping results demonstrate that strong performance on original arithmetic benchmarks does not guarantee robustness to controlled changes in the numeric instantiation of the same problem schema. The effect is most pronounced on GSM8K, where all completed model results show nontrivial conditional drops, and weakest on MAWPS and MultiArith, where retained attacks are solved at high rates by most models.

\section{Discussion}\label{sec:discussion}
The results show that strong benchmark performance on original arithmetic word problems does not necessarily imply robustness under schema-preserving numeric variation, but the size of this effect depends strongly on the dataset. Numeric remapping preserves the original reasoning program while changing the concrete quantities used to instantiate that program. In principle, a model that has learned the underlying arithmetic schema should transfer its reasoning from the original problem to the remapped version. In practice, the high-confidence attack results show a more nuanced pattern: GSM8K remains substantially affected by numeric remapping, while MAWPS and MultiArith exhibit much smaller drops.

The clearest robustness failures occur on GSM8K. Among the completed GSM8K evaluations, Gemma4 falls to \(74.18\%\) conditional attacked accuracy, DeepSeek-R1 reaches \(80.60\%\), and GPT-OSS reaches \(87.84\%\). These correspond to conditional drops of (25.82), (19.40), and (12.16) percentage points, respectively. These drops are meaningful because the final GSM8K attack set has been filtered through a strict high-confidence audit. The retained attacks are not arbitrary adversarial corruptions: they preserve the symbolic computation, use recomputed gold answers, and pass additional checks for surface consistency. The remaining failures therefore suggest that even valid schema-preserving numeric changes can disrupt model behavior on more complex arithmetic word problems.

At the same time, the MAWPS and MultiArith results show that numeric remapping is not uniformly difficult across arithmetic datasets. On MAWPS, completed model evaluations remain above \(98\%\) conditional attacked accuracy, with only small conditional drops. GPT-OSS reaches \(98.81\%\), Gemma4 reaches \(98.25\%\), and DeepSeek-R1 reaches \(96.62\%\). MultiArith shows a similar pattern: GPT-OSS answers all retained attacks correctly, while Gemma4 and DeepSeek-R1 remain at \(98.42\%\) and \(98.25\%\), respectively. These results suggest that numeric-remapping robustness depends not only on the model, but also on the structure of the benchmark. MAWPS and MultiArith problems are generally shorter, more direct, and more template-like than GSM8K problems. When the reasoning path is simple and explicitly signaled, changing the numbers appears much less disruptive. In contrast, GSM8K contains longer and more linguistically varied problems, creating more opportunities for both model reasoning failures and surface-realization complications.

The attack-set construction results reinforce this dataset-level difference. After post-hoc high-confidence auditing, the pipeline retains (732) attacks from (1{,}885) GSM8K source examples, compared with (1{,}597) from (1{,}700) MAWPS examples and (570) from (593) MultiArith examples. This lower GSM8K retention rate reflects the difficulty of producing high-confidence numeric remappings for longer and more varied word problems. In particular, surface realization can be subtle: a single symbolic variable may appear through multiple natural-language mentions, and failing to update all relevant mentions can produce an attacked question whose text no longer matches the recomputed answer. The additional audit step therefore plays an important role in separating valid robustness examples from candidate attacks with surface-level inconsistencies.

The high-confidence audit also affects how the results should be interpreted. The final reported attack sets are intentionally conservative: examples are retained only when the symbolic computation, remapped answer, and rendered surface text align under deterministic checks. This means the results should be read as performance on a smaller but more trustworthy set of numeric-remapping attacks, rather than as an estimate over every possible generated candidate. Under this stricter evaluation, the central finding is not that all arithmetic datasets are highly vulnerable to numeric remapping, but that robustness varies sharply by dataset: GSM8K shows persistent sensitivity, while MAWPS and MultiArith are largely stable under the retained transformations.

The computational results further emphasize that generalization-attack evaluation is both a modeling problem and an infrastructure problem. Numeric-remapping attack generation requires repeated large-model calls, symbolic checks, retries, surface edit-plan generation, deterministic rendering, and post-hoc auditing. The runtime comparison between the ZGX Nano and GH200 Supermicro shows that hardware substantially affects the feasibility of constructing attacked benchmarks at scale. Faster generation makes it possible to debug prompts, inspect invalid examples, regenerate attacks, rerun audits, and evaluate multiple target models. This matters because the final benchmark is not produced in a single pass: the pipeline must be iterated and audited carefully to avoid conflating model failures with attack-generation artifacts.

Overall, the numeric-remapping results support a refined version of the central claim of this paper: static benchmark accuracy is an incomplete measure of arithmetic reasoning robustness, especially for more linguistically complex word-problem datasets. A model may answer an original problem correctly while failing on a valid transformed version that preserves the same symbolic reasoning program. However, the results also show that robustness is not a single global property of a model. It depends on the interaction between model behavior, dataset structure, transformation type, and validation strictness. Numeric remapping is therefore useful not only as an attack, but also as a diagnostic tool for identifying where benchmark success reflects stable schema-level reasoning and where it remains sensitive to the particular numeric and linguistic instantiation of the problem.


\section{Limitations}\label{sec:limitations}
This work has several limitations. First, the empirical evaluation focuses on numeric remapping only. The broader taxonomy in Section~\ref{sec:attack} describes other possible schema-preserving attacks, including lexical paraphrasing, unit conversion, distractor insertion, relation substitution, question inversion, and question merging. However, these families are not evaluated experimentally in the present paper. Numeric remapping is a useful first case because it preserves the original reasoning program and permits direct recomputation of the attacked gold answer, but it does not capture the full space of reasoning failures that may arise under other transformations.

Second, the attacked evaluation is conditional on the source model's original correctness and on successful attack generation. We generate numeric-remapping attacks from examples answered correctly by GPT-OSS (120B), then retain only examples that pass the structured generation pipeline and post-hoc high-confidence audit. This design is appropriate for testing whether originally correct reasoning survives controlled transformation, but it means that attacked accuracy should not be interpreted as accuracy on a random sample of the original dataset. Instead, it measures robustness on the subset of examples that are source-model-correct, successfully transformable, and retained under the high-confidence audit.

Third, the final attack sets are intentionally conservative. The post-hoc audit filters out examples with possible surface mismatches, stale source spans, malformed rendered text, or disagreement between the rendered question and recomputed answer. This improves trust in the retained attacks, but it also reduces coverage, especially for GSM8K. As a result, the reported attacked accuracies should be interpreted as performance on high-confidence numeric-remapping attacks rather than on every candidate attack produced by the pipeline. Some valid attacks may be excluded because the audit is strict, while some subtle semantic issues may still remain despite passing automatic checks.

Fourth, attack validity is checked primarily through automatic validation rather than exhaustive manual verification. The pipeline verifies that symbolic expressions recover the original answer, that remapped values satisfy extracted constraints, that attacked answers are recomputed from the symbolic program, and that rendered questions reflect the remapped quantities. These checks substantially reduce invalid attacks, but they cannot guarantee perfect semantic validity in every case. Some transformations may pass structural checks while still introducing subtle natural-language inconsistencies. Conversely, some valid transformations may be rejected because the symbolic extractor, surface edit planner, or audit heuristic fails. Future work should include more systematic manual audits and stage-wise error analysis.

Fifth, the pipeline depends on a language model for structured generation stages, including schema inference, constraint extraction, remapping, and surface edit-plan generation. Although the outputs are checked before being retained, the quality and distribution of generated attacks may still depend on the generation model, prompting strategy, retry budget, and parsing rules. In this work, GPT-OSS (120B) is used as the attack-generation model. Using a different generation model could change the retention rate, the distribution of retained attacks, or the kinds of examples that pass validation.

Sixth, model-output parsing and completion introduce an additional source of uncertainty. Some evaluated models, especially reasoning-oriented models, may produce long outputs that are truncated, malformed, or difficult to parse into a final numeric answer. In these cases, an apparent failure may reflect an output-format or completion issue rather than an arithmetic reasoning error. Final accuracy estimates may still depend on parsing choices and recovery heuristics. Future evaluations should standardize output formats more tightly or use evaluation prompts that better separate reasoning traces from final answers.




Finally, this work is limited to arithmetic word-problem benchmarks. Arithmetic reasoning is a useful testbed because valid transformations can often be checked through symbolic recomputation, but robustness failures in other domains may require different representations and validation procedures. Extending schema-preserving attack generation to domains such as code generation, scientific reasoning, logical inference, or multi-hop question answering remains an important direction for future work.

\section{Acknowledgment}
This work was supported in part by the National Science Foundation under Award No. 2509828, “NRT: Building Responsible AI Researchers: Advancing Research and Innovation at the Intersection of AI and Human Actors,” and by the Supermicro NVIDIA Grace Enablement Evaluation Program. The authors gratefully acknowledge NVIDIA and Supermicro for providing access to a Supermicro ARS-111GL-NHR server equipped with an NVIDIA GH200 Grace Hopper Superchip, which supported the evaluation and research activities reported in this paper.

\section{Conclusion}\label{sec:conclusion}
This paper introduced numeric remapping as a schema-preserving generalization attack for arithmetic word problems. The central idea is to transform a problem by changing its concrete quantities while preserving the original reasoning program. Because the symbolic structure is preserved, the attacked gold answer can be recomputed directly rather than guessed or manually assigned. Numeric remapping therefore provides a controlled way to test whether models that solve original benchmark problems remain correct on valid variants of those same problems.

We presented a structured pipeline for generating numeric-remapping attacks. The pipeline infers a symbolic representation of the original problem, extracts constraints on editable quantities, generates new numeric assignments, recomputes the attacked answer, and realizes the transformed question through an LLM-generated surface edit plan that is applied deterministically. In addition to stage-wise validation, we apply a post-hoc high-confidence audit to ensure that retained attacks align the rendered question, remapped quantities, symbolic program, and recomputed gold answer. This conservative filtering reduces coverage, especially on GSM8K, but produces a more trustworthy attacked benchmark.

Empirically, the results show that robustness under numeric remapping varies substantially by dataset. GSM8K exhibits the clearest degradation: among completed evaluations, Gemma4, DeepSeek-R1, and GPT-OSS all lose conditional accuracy under high-confidence numeric remapping, with GPT-OSS remaining the strongest completed GSM8K model at \(87.84\%\) attacked accuracy. In contrast, MAWPS and MultiArith show much smaller drops, with model evaluations remaining near or above \(96\%\) attacked accuracy. These findings suggest that numeric-remapping robustness is not a single global property of a model, but depends on the interaction between model behavior, dataset structure, and transformation difficulty.

More broadly, this work demonstrates that schema-preserving attacks can provide a practical diagnostic for model generalization. Static benchmark accuracy alone cannot show whether a model has learned a stable reasoning procedure or has succeeded on a particular numeric and linguistic instantiation of that procedure. Numeric remapping offers one automatically checkable starting point: the reasoning program is preserved, the values change, and the gold answer is recomputed. Future work can extend this framework to additional attack families, richer validation methods, more complete manual audits, and broader reasoning domains beyond arithmetic word problems.


\printbibliography

\section{Appendix}\label{sec:appendix}
\appendix

\section{Additional Numeric-Remapping Examples}
\label{app:numeric_examples}

Table~\ref{tab:additional_numeric_examples} provides additional examples of valid numeric-remapping attacks. Each attacked problem changes the concrete quantities in the original problem while preserving the underlying reasoning program and recomputing the gold answer from the transformed symbolic representation.

\begin{table}[htbp]
\centering
\caption{Additional examples of numeric-remapping attacks. Examples 1, 2, and 3 are drawn from GSM8K, MAWPS, and MultiArith, respectively.}
\label{tab:additional_numeric_examples}
\setlength{\tabcolsep}{4pt}
\begin{tabular}{p{0.28\linewidth}p{0.28\linewidth}p{0.18\linewidth}p{0.18\linewidth}}
\toprule
\textbf{Original question} & \textbf{Attacked question} & \textbf{Original answer} & \textbf{Attacked answer} \\
\midrule
Nathan is buying decorations for his wedding reception. The reception hall will have 20 tables. Each table needs a linen tablecloth (\$25 to rent), 4 place settings (\$10 each to rent), and a centerpiece. Each centerpiece will have 10 roses (\$5 each) and 15 lilies (\$4 each). How much will the decorations cost? & Nathan is buying decorations for his wedding reception. The reception hall will have 25 tables. Each table needs a linen tablecloth (\$30 to rent), \$3 place settings (\$12 each to rent), and a centerpiece. Each centerpiece will have 8 roses (\$6 each) and 12 lilies (\$5 each). How much will the decorations cost? & 3500 & 4350 \\
Mrs. Hilt measured the distance from her desk to the water fountain. It was 30 feet. How many feet will Mrs. Hilt walk on her trips to the fountain if she goes to the water fountain 4 times today? & Mrs. Hilt measured the distance from her desk to the water fountain. It was 25 feet. How many feet will Mrs. Hilt walk on her trips to the fountain if she goes to the water fountain 5 times today? & 120 & 125 \\
Vanessa and her mom were picking carrots from their garden. Vanessa picked 17 and her mother picked 14. If only 24 of the carrots were good, how many bad carrots did they have? & Vanessa and her mom were picking carrots from their garden. Vanessa picked 18 and her mother picked 15. If only 20 of the carrots were good, how many bad carrots did they have? & 7 & 13 \\
\bottomrule
\end{tabular}
\end{table}

\section{Validation and Filtering Criteria}
\label{app:validation}

The numeric-remapping pipeline uses stage-wise validation to filter invalid examples before they enter the final attacked benchmark. This appendix summarizes the main checks used during attack construction.

\paragraph{Schema inference.}
An example passes schema inference if the generated symbolic expression can be parsed and executed, all free symbols in the expression align with extracted variable definitions, and substituting the original variable values into the expression reproduces the original gold answer.

\paragraph{Constraint extraction.}
An example passes constraint extraction if each extracted variable contains the required fields, including its original value, source text, replaceability status, and relevant numeric or semantic constraints. Variables marked as fixed must remain unchanged during remapping, while variables marked as replaceable must receive valid transformed values.

\paragraph{Numeric remapping.}
An example passes numeric remapping if all editable variables receive replacement values, fixed variables remain unchanged, and the new values satisfy the extracted constraints. The transformed assignment must be complete enough to evaluate the symbolic expression and compute a new attacked gold answer.

\paragraph{Surface realization.}
An example passes surface realization if the rendered attacked question is aligned with the transformed variable assignments and remains a coherent standalone word problem. The pipeline rewrites only quantities that can be tied back to visible source text in the original question. The final attacked example is retained only if the rendered question and recomputed answer remain consistent with the transformed symbolic representation.

\section{Structured Output Templates}
\label{app:structured_templates}

The pipeline represents intermediate artifacts using structured fields rather than unrestricted free-form text. The exact implementation may vary, but the following templates summarize the expected information passed between stages.

\subsection{Symbolic Artifact}

\begin{verbatim}
{
  "expression": "final = ...",
  "variables": {
    "variable_name": {
      "source_text": "...",
      "original_value": ...,
      "meaning": "..."
    }
  },
  "verified": true
}
\end{verbatim}

\subsection{Constraint Artifact}

\begin{verbatim}
{
  "inputs": [
    {
      "name": "variable_name",
      "source_text": "...",
      "value": ...,
      "replaceable": true,
      "numeric_rules": [...],
      "semantic_role": "..."
    }
  ]
}
\end{verbatim}

\subsection{Remapping Artifact}

\begin{verbatim}
{
  "remap": {
    "variable_name": {
      "value": ...,
      "replacement_text": "..."
    }
  }
}
\end{verbatim}

\section{Stage Checks for Numeric Remapping}
\label{app:stage_checks}

\begin{algorithm}[htbp]
\caption{Stage Checks for Numeric Remapping}
\label{alg:numeric_remapping_checks}
\begin{algorithmic}[1]

\Function{ValidSchema}{$S,x,y$}
    \State Check that $S$ contains a symbolic expression and variable definitions
    \State Check that all free symbols in the expression appear in the variable definitions
    \State Substitute original variable values into the symbolic expression
    \State \Return the evaluated expression equals the original gold answer $y$
\EndFunction

\Function{ValidConstraints}{$C,S$}
    \State Check that every variable in $S$ has a corresponding constraint entry in $C$
    \State Check that each entry specifies source text, original value, replaceability, and semantic role
    \State Check that numeric and semantic rules are well formed
    \State \Return all required constraint fields are present and valid
\EndFunction

\Function{ValidRemap}{$R,S,C$}
    \State Check that each editable variable receives a replacement value
    \State Check that fixed variables remain unchanged
    \State Check that replacement values satisfy extracted numeric and semantic constraints
    \State Check that evaluating the symbolic expression under $R$ produces a valid numeric answer
    \State \Return the remapping is complete, computable, and constraint-satisfying
\EndFunction

\Function{ValidEditPlan}{$E,x,R$}
    \State Check that $E$ contains structured edit actions rather than a free-form rewritten question
    \State Check that each edit specifies a variable, an original text span, and replacement text
    \State Check that each edited variable appears in the remapping $R$
    \State Check that every original text span appears exactly in the source question $x$
    \State Check that proposed edit spans are non-overlapping
    \State Check that all editable remapped variables are represented by at least one edit
    \State \Return the edit plan can be safely applied to $x$
\EndFunction

\Function{ApplySurfaceEdits}{$x,E$}
    \State Sort edit actions in $E$ by decreasing character position in $x$
    \State Initialize $x' \gets x$
    \ForAll{edit actions $e \in E$ in sorted order}
        \State Replace the exact source span for $e$ with its replacement text
    \EndFor
    \State \Return rendered attacked question $x'$
\EndFunction

\Function{ValidRenderedAttack}{$x',y',S,C,R,E$}
    \State Check that remapped quantities and replacement text appear in the rendered question $x'$
    \State Check that stale source spans targeted by $E$ do not remain in $x'$
    \State Check that no malformed surface patterns are introduced by applying $E$
    \State Check that unchanged quantities remain aligned with the original problem
    \State Check that $y'$ is produced by evaluating the symbolic expression in $S$ under $R$
    \State \Return $(x',y')$ is coherent, solvable, and correctly labeled
\EndFunction

\end{algorithmic}
\end{algorithm}

\end{document}